\documentstyle[eqsecnum,aps,epsfig]{revtex}
\def\beq{\begin{eqnarray}} \def\eeq{\end{eqnarray}}

\newcommand\be{\begin{equation}}

\def\mev{\,{\rm MeV}}
\def\gev{\,{\rm GeV}}

\renewcommand{\d}{\rm{d}}
\renewcommand{\u}{\rm{u}}
\newcommand{\q}{\rm{q}}
\newcommand{\s}{\rm{s}}

\newcommand{\dbar}{\overline{\rm{d}}}
\newcommand{\ubar}{\overline{\rm{u}}}
\newcommand{\sbar}{\overline{\rm{s}}}

\def\qbq{{\rm q}\overline{\rm q}}
\def\sbs{{\rm s}\overline{\rm s}}

\begin{document}
\preprint{\parbox{5cm}{Wuppertal WUB 98-7\\[-0.5em]hep-ph/9805294}}
\title{Interpolation Formulas for the $\eta\gamma$ and
$\eta'\gamma$ Transition Form Factors}
\author{ Th.\ Feldmann\thanks{Supported by {\it Deutsche
      Forschungsgemeinschaft}}  and P.\ Kroll}
\address{ Fachbereich Physik, Universit\"at Wuppertal,
D-42097 Wuppertal, Germany}
\date{to be published in {\it Physical Review }\/ D}
\maketitle
\begin{abstract}
The new CLEO and LEP data \cite{CLEO97,L397}
on the  $\eta\gamma$ and
$\eta'\gamma$ transition form factors have renewed the
interest in simple interpolation formulas, valid at any
value of momentum transfer.
We are going to show that 
recent theoretical and phenomenological
results on $\eta$--$\eta'$ mixing \cite{Leutwyler97,FeKr97b,FeKrSt98}
lead to two-pole forms, where each pole term resembles 
the Brodsky/Lepage interpolation
formula for the $\pi\gamma$ case
\cite{BLint} and depends on the mixing and decay parameters in
a simple fashion.
The parameters, entering the occasionally used 
one-pole formulas, on the other hand, cannot be interpreted  
theoretically in a simple way.
\vskip3em

\begin{center}
{\tt Copyright 1998 by The American Physical Society.}
\end{center}

\vskip3em

\end{abstract}
\section{Introduction}

In 1995 the CLEO collaboration has presented their preliminary data on
pseudoscalar meson-photon transition form factors (i.e.\
on the processes $\gamma\gamma^* \to P$, $P=\pi,\eta,\eta',\ldots$)
at large space-like momentum transfer, $Q^2$, for the first time
\cite{CLEO95}. One of
the exciting aspects of these form factors is that they
possess well-established asymptotic limits \cite{WaZe73,BrLe80},
which are related to mesonic decay constants only and 
behave as $1/Q^2$. 
On the other hand, at vanishing momentum transfer, $Q^2=0$,
the values of the form factors are fixed by PCAC and the chiral anomaly
and appear to be proportional to the inverse of the mesonic 
decay constants.
The investigation of the transition form factors between the two
limiting cases provides important information on the wave functions
of the pseudoscalar mesons.

In case of the $\pi^0$, where the decay constant is known from
the leptonic $\pi$ decay, Brodsky and Lepage (BL) \cite{BLint} 
proposed a simple
parameter-free formula which interpolates 
between the PCAC value at $Q^2=0$ and
the QCD prediction for $Q^2 \to \infty$
\beq
F_{P\gamma}^{BL}(Q^2) & = &
\frac{6 \, C_P \, f_P}{Q^2 + 4\pi^2 f^2_P}
 \ , \quad (P=\pi)
\label{BLint}
\eeq
($C_\pi = 1/(3 \sqrt2)$ represents a charge factor). 
This formula works fairly well
although it is not perfect, and may be used as an
estimate of the $\pi\gamma$ transition form factor. 
Since the $x$-dependence (where $x$ is
the usual momentum fraction the quark carries) of the pion's wave
function seems to be close to that of the asymptotic form
\cite{KrRa96,MuRa97}, the phenomenological success of the BL 
interpolation formula is not a surprise. 
Note that the vector meson
dominance model also provides a formula like (\ref{BLint}). It 
is very close to the BL interpolation formula numerically since
$M_\rho$ approximately equals $2\pi f_\pi$. Even the pole
residue agrees rather well with $6 \,C_\pi\, f_\pi$ numerically.
  
In case of the $\eta\gamma$ and $\eta'\gamma$ the situation is
more complicated due to the $\eta$--$\eta'$ mixing
which has to be taken into account
properly.
 $\eta - \eta'$ mixing has been the subject of considerable interest
and being examined in many investigations,
e.g.\ \cite{Fritzsch,Isgur,Gilman:1987ax} and references therein.
Recent investigations 
\cite{Leutwyler97,FeKr97b,FeKrSt98} have shown, that
the decay constants in the $\eta$-$\eta'$ system are not adequately
described in the octet-singlet scheme  with one and the same angle
for the particle state mixing and the parameterization of the
decay constants. 
As has become clear very recently \cite{FeKrSt98},
a mixing-scheme where one takes as the basis
the states according to their quark flavor decomposition
is better suited for the description
of the $\eta$ and $\eta'$ decay constants and meets all
theoretical and phenomenological requirements.
It also allows for
proper definitions of the desired generalizations of (\ref{BLint})
to the $\eta$ and $\eta'$ cases.
We define the quark basis by the two states $\eta_{\q}$ and $\eta_{\s}$
composed of the valence quarks $\qbq=(\u\ubar+
\d\dbar)/\sqrt{2}$ and $\sbs$ (and corresponding higher Fock
states), respectively. These states are related
to the physical states by the transformation
\be
 \left (\matrix{\eta \cr \eta'}\right )\,=\, U(\phi)\;
                      \left (\matrix{\eta_{\q} \cr \eta_{\s}}\right ) ,
\label{qsb}
\qquad
    U(\phi)\,=\,\left(\matrix{\cos{\phi} & -\sin{\phi} \cr
                                \sin{\phi} &
    \phantom{-}\cos{\phi}} \right) .
\end{equation}
The four decay constants of the $\eta$ and
$\eta'$ mesons, which are defined by
\be
\langle 0 | J_{\mu5}^i | P \rangle = \imath \,
        f_P^i \, p_\mu ,  \qquad ( i,j=\q,\s,\ P=\eta,\eta' ) ,
\label{decqs}
\end{equation}
can be expressed in terms of three parameters: the
mixing angle $\phi$ and two decay constants $f_{\q}$ and $f_{\s}$,
\beq
\left(  \matrix{f_{\eta}^{\q} & f_{\eta}^{\s} \cr
          f_{\eta'}^{\q} & f_{\eta'}^{\s}} \right)
&=& U(\phi) \, \left(\matrix{f_{\q}& 0 \cr 0 & f_{\s}}\right) .
\label{fdec}
\eeq
Here, the weak axial-vector currents are defined by $J^{\q}_{\mu 5}= (\ubar
\gamma_{\mu} \gamma_5 \u + \dbar\gamma_{\mu} \gamma_5 \d )/\sqrt{2}$
and $J^{\s}_{\mu 5}= \sbar\gamma_{\mu} \gamma_5 \s$.
We repeat that in the conventional octet-singlet
scheme the decay constants do NOT follow the pattern of state mixing, i.e.\
the angles appearing in the state mixing and in the parameterization
of the decay constants are not identical
\cite{Leutwyler97,FeKr97b,FeKrSt98,KiPe93}. 
On exploiting the divergences of the axial-vector currents, the basic
mixing parameters $\phi$, $f_q$ and $f_s$ are fixed to first order
in the flavor symmetry breaking. 
In order to take into account higher order corrections the
mixing parameters can be determined phenomenologically.
In \cite{FeKrSt98} the following values have been shown to give
a consistent description of the available experimental data and
to be in line with the constraints from the
axial vector anomaly
\beq
&&\phi = 39.3^\circ \pm
1.0^\circ, \quad f_{\q} =(1.07 \pm 0.02) \,  f_\pi, \quad
f_{\s} = (1.34 \pm 0.06) \, f_\pi.
\label{fit}
\eeq
These phenomenological values agree within $10\%$ with the
theoretical values \cite{FeKrSt98} and with the results
from chiral perturbation theory \cite{Leutwyler97}.
We will use the values given in (\ref{fit})
for the following discussion of the
$\eta\gamma$ and $\eta'\gamma$ transition form factors. In
\cite{FeKrSt98} it has already been found that they
give a perfect description of the experimental
$\eta\gamma$ and $\eta'\gamma$
form factor data above $Q^2 = 1$~GeV$^2$ if the form factors are calculated
within the modified hard-scattering approach \cite{FeKr97b} and
if wave functions similar to the pion one are used.

\section{Generalized BL interpolation formulas}
%

Within the $\qbq$--$\sbs$ mixing scheme the following generalization
of (\ref{BLint}) seems to be most natural
($P=\eta,\eta'$)
\beq
F_{P\gamma}^{BL}(Q^2) & = &
\frac{6\,C_{\q}\,f_P^{\q}}{Q^2 + 4 \pi^2 \, f_{\q}^2}
+\frac{6\,C_{\s}\,f_P^{\s}}{Q^2 + 4 \pi^2 \, f_{\s}^2} \ ,
\label{etainterpolqs}
\eeq
where the charge factors are $C_{\q}=5/9\sqrt2,\ C_{\s}=1/9$.
At $Q^2=0$ it reproduces the PCAC formula (see e.g.\ \cite{FeKrSt98})
which enters the two-photon widths of the $\eta$ and $\eta'$ mesons.
For $Q^2\to\infty$ QCD predicts the value of the form factors in
terms of the decay constants $f_P^i$ defined by (\ref{decqs}).
The latter represent
wave functions of the lowest parton Fock states, $\qbq$ and $\sbs$,
contributing to the $\eta$ and $\eta'$ mesons, taken at zero spatial
separation, which appear in the perturbative calculation of
the form factor (see e.g.\ \cite{FeKr97b}). 
The numerical factors in (\ref{etainterpolqs}) are
chosen in such a way that the asymptotic behavior, as
predicted by QCD, is reproduced.

\begin{figure}[hbtp]
\begin{center}
{\psfig{file=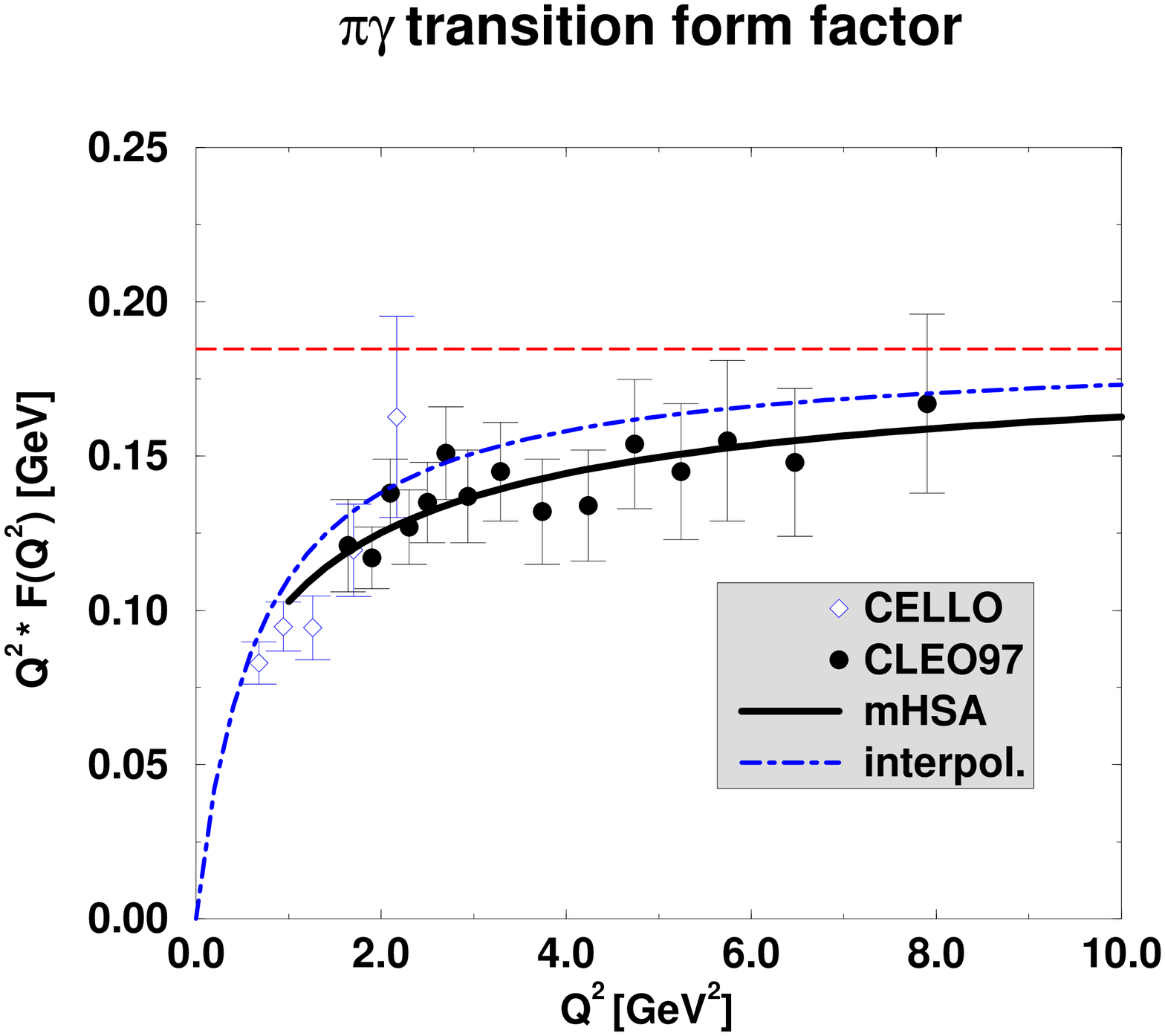, width=4.3cm,angle = -90}} \
{\psfig{file=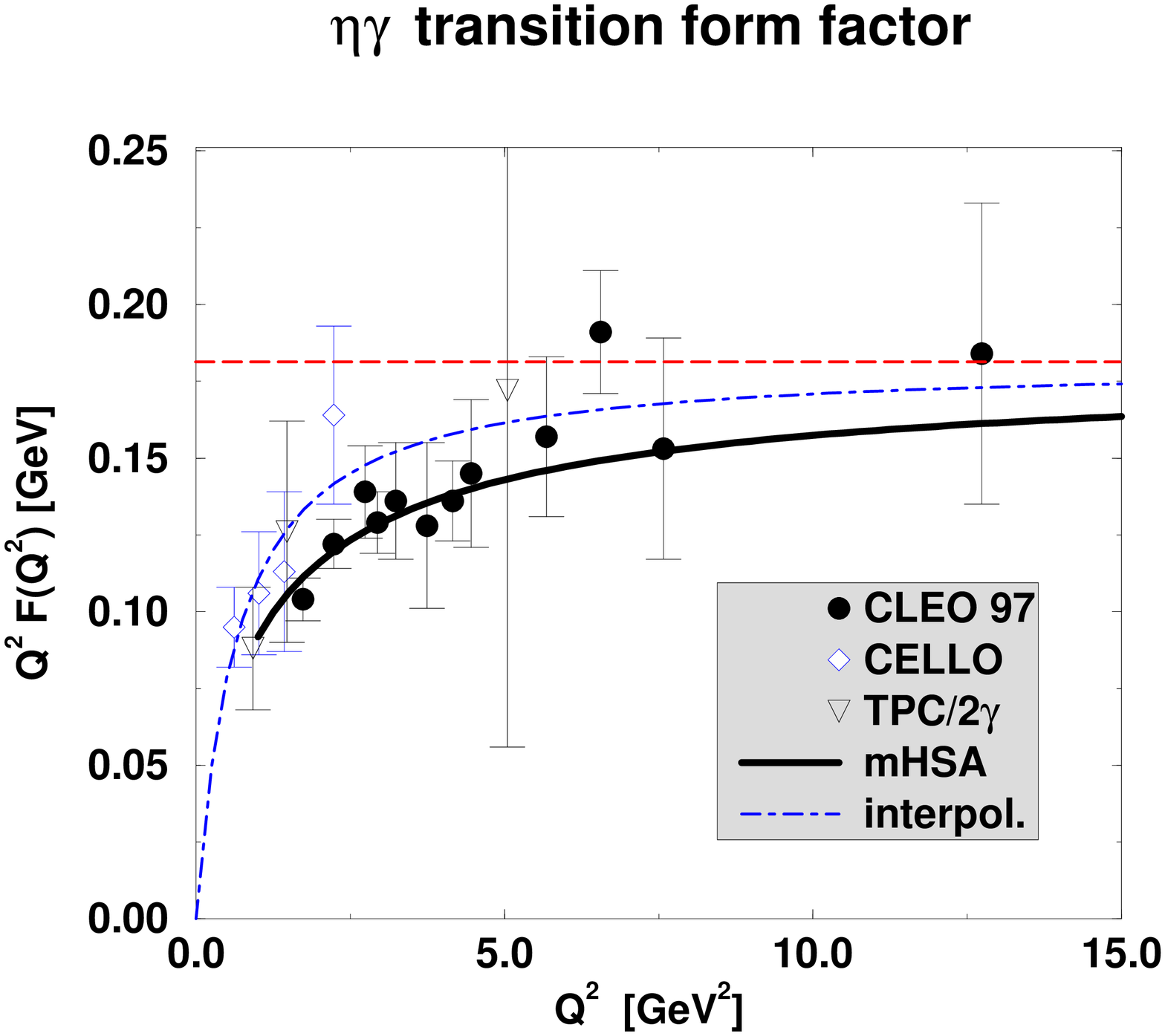, width=4.3cm,angle = -90}} \
\hskip-.5cm
{\psfig{file=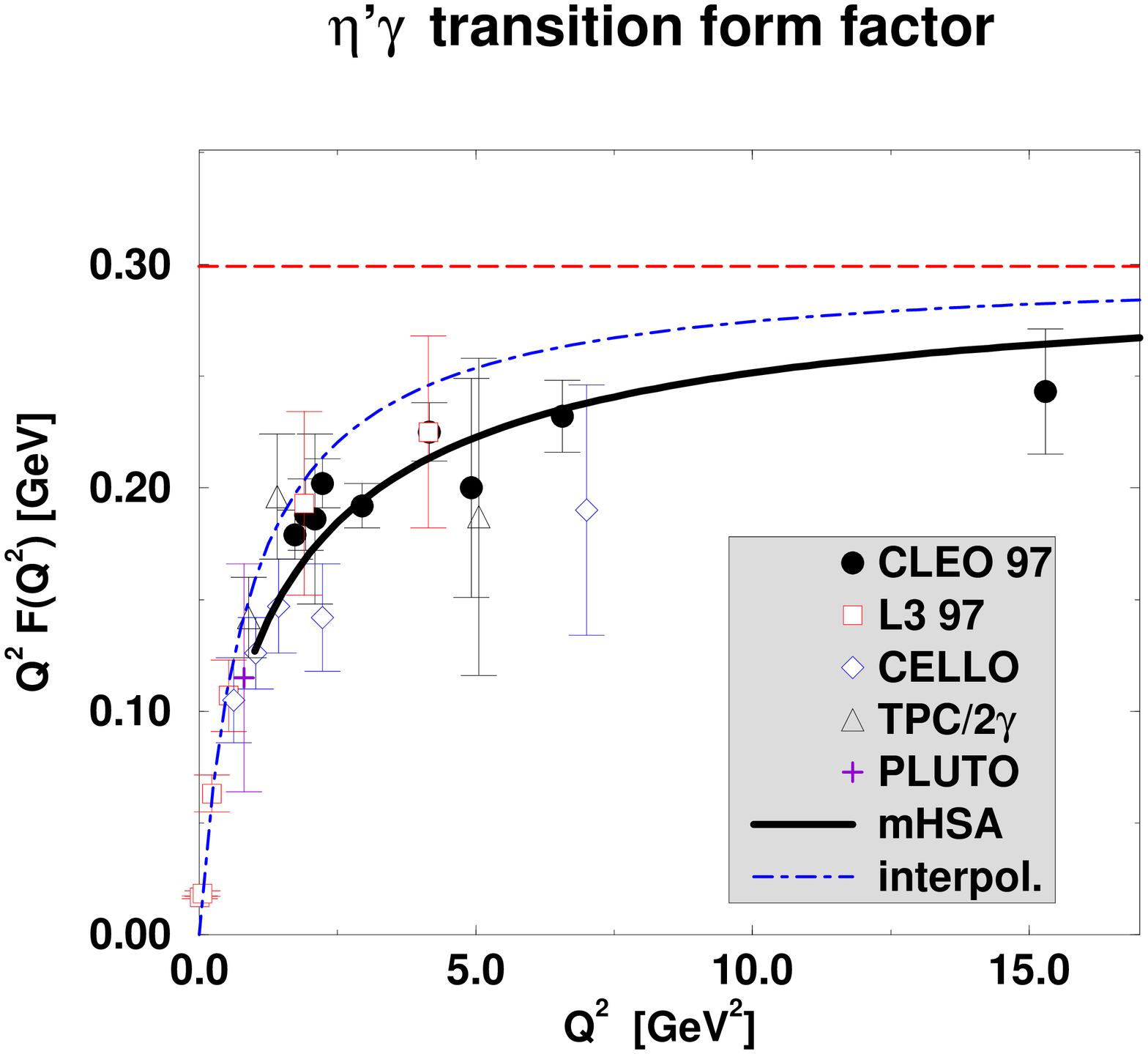, width=4.3cm,angle = -90}} 
\end{center}
\caption{Results for the $\pi\gamma$, $\eta\gamma$ and
$\eta'\gamma$ transition form factors calculated
using the modified hard scattering approach
with asymptotic wave functions,
and with the interpolation formulas (\ref{BLint},\ref{etainterpolqs}).
The straight lines indicate the asymptotic limit.
The parameter values are $f_\pi=131$~\mev,
$\phi_P=39.3^\circ$, $f_{\q}=1.07 f_\pi$,
$f_{\s}=1.34 \, f_\pi$.
Data are taken from \protect\cite{CLEO97,L397,CELLO91,TPC90,PLUTO}. }
\label{Interpolfig}
\end{figure}

Comparison with experiment (see Fig.~\ref{Interpolfig}) reveals 
that the interpolation
formulas (\ref{BLint}) and  (\ref{etainterpolqs}) do not work badly.
However, they should still be 
used with care: For momentum transfer of the order of a few GeV$^2$ the
predicted curvatures of the form factors are stronger than
the present data exhibit. 
On the other hand, the predictions obtained within
the modified hard scattering approach are in perfect agreement with
experiment \cite{FeKr97b,FeKrSt98,KrRa96} above 1~GeV$^2$.

Our generalization of the BL interpolation formula (\ref{etainterpolqs})
parallels the vector meson dominance model. This is so because
the $\omega$ meson is nearly a pure $\qbq$ state and mass degenerated
with the $\rho$, while the $\phi$ meson is almost a pure $\sbs$ state.
In the vector meson dominance ansatz 
\beq
F_{P\gamma}^{VMD}(Q^2) = \sum_{V=\rho,\omega,\phi} 
\frac{g_{PV\gamma}\, M_V^2}{f_V} \, \frac{1}{Q^2+M_V^2}
\eeq
one can effectively combine the first two terms into one because
$M_\rho\simeq M_\omega$. This term can then be identified with the first
term in (\ref{etainterpolqs}), since $M_{\rho,\omega} \simeq
2\pi\,f_{\q}$.
The remaining contribution of the $\phi$ meson corresponds to
the second term in (\ref{etainterpolqs}) since
$M_\phi \simeq 2 \pi f_{\s}$.
Moreover, as a detailed analysis reveals, the corresponding
residues in the generalized BL interpolation formula and in the vector
meson dominance ansatz approximately agree as well (for values
of $g_{PV\gamma}$ and $f_V$ see \cite{Dumbrajs:1983jd,BaFrTy95}).

Occasionally, one-pole BL interpolation formulas, like (\ref{BLint}),
are used for the $\eta$ and $\eta'$ cases
\cite{CLEO97,L397,CELLO91,TPC90} (see also the tables of particle data
\cite{PDG96}).
The corresponding parameters, $C_P$ and $f_P$ ($P=\eta,\eta'$),
however, have no simple theoretical interpretation. In particular
$f_P$, misleadingly called {\lq}decay constant{\rq}, does not
represent a vacuum-particle matrix element of an appropriate
axial-vector current and is, therefore, not related to basic decay
properties of the $\eta$ and $\eta'$ mesons in a simple way.
The requirement that the one-pole formulas interpolate between
the PCAC results at $Q^2=0$ and the asymptotic perturbative QCD
predictions, relates the parameters $C_P$ and $f_P$ to the decay
constants $f_P^i$. Using the $\qbq$--$\sbs$ mixing scheme one
finds
\beq
C_P &=& \sqrt{ (C_q \, f_P^q + C_s \, f_P^s) \,
               ( C_q \, f_P^q/f_q^2 +C_s \, f_P^s/f_s^2)}
\ , \nonumber \\[0.3em]
f_P &=& \sqrt{ \frac{C_q \, f_P^q + C_s \, f_P^s}{
               C_q \, f_P^q/f_q^2 +C_s \, f_P^s/f_s^2}}
\eeq
Despite of the fact that the parameters of the one-pole
formula have a much more complicated interpretation than
those  of (\ref{etainterpolqs}), the one-pole formula is
of similar quality. Using, for instance, the values (\ref{fit})
for the basic mixing parameters to estimate $C_P$ and $f_P$,
one obtains results from the one-pole formula which are almost
indistinguishable from what is found from (\ref{etainterpolqs}).

Amettler et al.\ \cite{Bijnensetal92} also proposed a
one-pole BL interpolation formula for the $\eta\gamma$ and
$\eta'\gamma$ transition form factors. Since they implicitly assumed
$C_P=C_\pi$, their formula only matches the PCAC result at
$Q^2=0$
while it fails asymptotically and, hence, at large $Q^2$.

\section{Summary}

We have proposed BL interpolation formulas for
the $\eta\gamma$ and $\eta'\gamma$ transition form factors which 
resemble the BL formula for the $\pi\gamma$
case. The formulas (\ref{etainterpolqs}) 
are derived within the new mixing scheme advocated for in 
\cite{FeKrSt98}, where
the $\eta$ and $\eta'$ mesons are expressed as linear combinations
of orthogonal states $\eta_q$ and $\eta_s$.
Only in this mixing scheme 
the decay constants $f_P^i$ ($i=q,s$, $P=\eta,\eta'$)
can be described in terms of two basic decay constants $f_{\q}$, $f_{\s}$ 
and the state mixing angle $\phi$. This is a necessary requirement
for the BL interpolation formula to be of the desired simple form.
It is then rather similar to 
the vector meson dominance model, since
the vector mesons show nearly ideal mixing, i.e.\ they distinguish
between the $\sbs$ and $\qbq$ components in a natural way.
Thus, a two-pole ansatz for the BL interpolation formula
appears to be very natural. Its other advantage is that the
basic mixing parameters, $f_q,\, f_s,\, \phi$, enter in a simple
fashion.
With the values (\ref{fit}) for the basic mixing parameters
the interpolation formulas
(\ref{etainterpolqs})
are in rough agreement with the experimental data 
\cite{CLEO97,L397,CELLO91,TPC90,PLUTO}.
Above
$Q^2=1~\gev^2$ the data are much better reproduced within a perturbative
approach \cite{FeKr97b,FeKrSt98}. Despite of this,
the interpolation formula 
is illustrative and may be 
used for quick estimates of the form factors. 

It is also possible to specify one-pole BL interpolation
formulas for the $\eta\gamma$ and $\eta'\gamma$ transition
form factors. However, the parameters appearing in the one-pole
formulas, have no simple theoretical interpretation; they depend
on the basic mixing parameters in a very complicated way.
In particular, the parameters $f_P$ cannot be interpreted as
{\lq}decay constants{\rq} as is done occasionally.

\end{document}